\documentclass[aps,twocolumn,nofootinbib,preprintnumbers,groupedaddress,letterpaper]{revtex4-2}

\usepackage[dvips]{color} \usepackage[normalem]{ulem} \usepackage{amsmath}
\usepackage{enumerate} \usepackage{amsfonts} \usepackage{epsfig}
\usepackage{yfonts}
\usepackage{slashed}
\usepackage{xcolor}

\usepackage{graphicx} \usepackage{bm} \usepackage{subfigure}

\newcommand\comment[1]{}

\newcommand\om{\omega}  \newcommand\ov{\over }

\def\({\left(} \def\){\right)} \def\[{\left[} \def\]{\right]} \def\<{\langle}
\def\>{\rangle}

\newcommand\half{{\ensuremath{\frac{1}{2}}}}

\newcommand{\be}{\begin{equation}} \newcommand{\ee}{\end{equation}}
\newcommand{\bea}{\begin{eqnarray}} \newcommand{\eea}{\end{eqnarray}}
\newcommand{\bwt}{\begin{widetext}} \newcommand{\ewt}{\end{widetext}}
 \newcommand{\bi}{\begin{itemize}}
\newcommand{\ei}{\end{itemize}} \newcommand{\ben}{\begin{enumerate}}
\newcommand{\een}{\end{enumerate}} \newcommand{\bca}{\begin{cases}}
\newcommand{\eca}{\end{cases}} \newcommand{\bln}{\begin{align}}
\newcommand{\eln}{\end{align}} \newcommand{\bst}{\begin{split}}
\newcommand{\est}{\end{split}}

\def\cS{{\cal S}}

\def\uuv{{\underline{v}}}
\def\uur{{\underline{r}}}

\def\uui{{\underline{i}}}

\def\grr{{\Gamma^{\uur}}}
\def\gvv{{\Gamma^{\uuv}}}

\def\gvi{{\Gamma^{\underline{vi}}}}

\def\g2{{\Gamma^{(2)}}}

\def\ief{{ingoing Eddington-Finkelstein coordinates}}

\def\ief{{ingoing Eddington-Finkelstein coordinates}}

\newcommand{\rz}[1]{\psi^{(#1,#1)}_{r, 0}}
\newcommand{\vz}[1]{\psi^{(#1,#1)}_{v, 0}}
\newcommand{\rg}[1]{\psi^{(#1,#1)}_{r}}
\newcommand{\vg}[1]{\psi^{(#1,#1)}_{v}}

\allowdisplaybreaks

\begin{document}
\preprint{QMUL-PH-21-01}

\title{Pole-skipping and Rarita-Schwinger fields}

\author{Nejc \v{C}eplak$^1$}
\email{\href{mailto:nejc.ceplak@ipht.fr}{nejc.ceplak@ipht.fr}}
\author{David Vegh$^2$}
\email{\href{mailto:d.vegh@qmul.ac.uk}{d.vegh@qmul.ac.uk}}

\affiliation{\it $^1$ Institut de Physique Th\'eorique,
Universit\'e Paris Saclay,\\
CEA, CNRS, F-91191 Gif sur Yvette, France}
\affiliation{\it $^2$ Centre for Research in String Theory, School of Physics and Astronomy \\
Queen Mary University of London, 327 Mile End Road, London E1 4NS, UK  }

\begin{abstract}

\noindent
In this note we analyse the equations of motion of a minimally coupled Rarita-Schwinger field near the horizon of  an anti-de Sitter-Schwarzschild geometry.
We find that at special complex values of the frequency and momentum  there exist two independent regular solutions that are ingoing at the horizon.
These special points in Fourier space are associated with the `pole-skipping' phenomenon in thermal two-point functions of operators that are holographically dual to the bulk fields.
We find that the leading pole-skipping point is located at a positive imaginary frequency with the distance from the origin being equal to half of the Lyapunov exponent for maximally chaotic theories.
\end{abstract}

\maketitle

\section{Introduction}
\label{sec:intro}

Retarded Green's functions are one of the main objects of interest in field theories as they encode the response of a system in equilibrium to small perturbations.
In a theory with a dual gravitational description \cite{Maldacena:1997re} one can calculate such correlators at finite temperature by studying the equations of motion of  bulk fields in a black hole geometry which  asymptotes to anti-de Sitter (AdS) spacetime \cite{Gubser:1998bc, Witten:1998qj}.
In Lorentzian signature, one is prescribed to take the solution that is ingoing at the horizon \cite{Son:2002sd}
which uniquely determines the retarded Green's function of the dual boundary operators.
Finding the precise form of the correlation function is generically difficult as the full solutions to the to the bulk equations of motion  depend on the details of the background geometry.

However, some features of the Green's functions are encoded in the behaviour of the bulk fields in the near-horizon region of the geometry.
The most prominent example is the hydrodynamic description of holographic theories, described by the low-frequency and low-momentum limit (see for example \cite{Kovtun:2012rj, Glorioso:2018wxw}), where one finds that the radial evolution of the bulk fields becomes trivial \cite{Iqbal:2008by} and leads to universal results such as the ratio of the shear viscosity and the entropy density \cite{Kovtun:2004de}.

Interestingly, the horizon region also contains information about the correlator away from the origin of Fourier space.
Namely, at certain imaginary values of the frequency and (generically) complex values of the momentum, all solutions to the equations of motion become ingoing at the horizon.
This phenomenon was called `pole-skipping' as at such points in momentum space the associated boundary Green's function effectively skips a pole: at these locations a zero and a pole of the correlator coincide.

\begin{figure}[h!]
\centering
\includegraphics[width =0.7\columnwidth ]{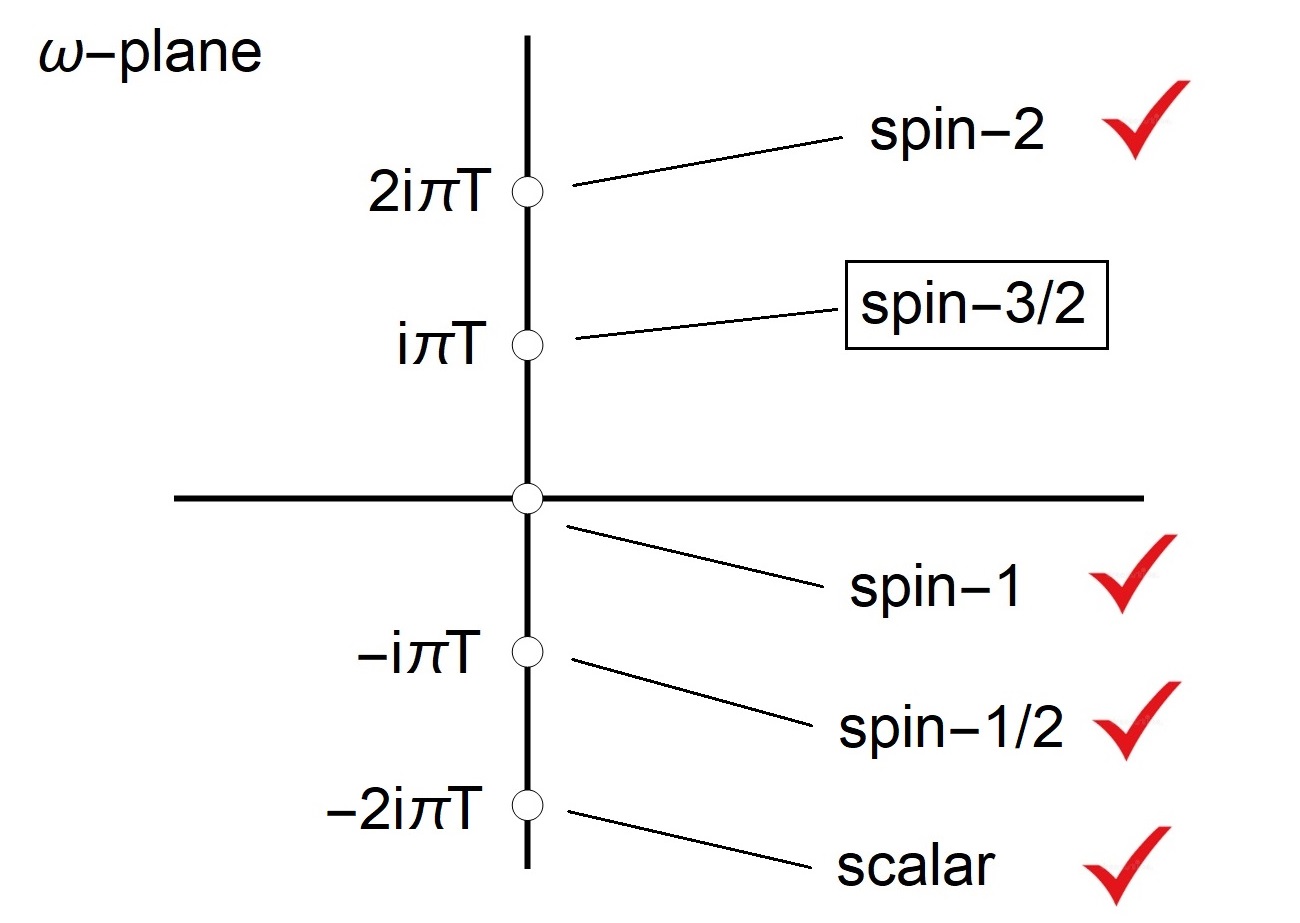}
\caption{\label{figorder} Location of the leading pole-skipping point on the frequency plane for fields of various spin. This paper focuses on the spin-${3\ov 2}$ case.}
\end{figure}

Pole-skipping was initially observed in the energy density correlator which is obtained by solving the linearised Einstein's equations in the bulk.
 One finds that the leading pole-skipping point, that is the pole-skipping point at the most positive imaginary frequency (see FIG.~\ref{figorder}),  is located at \cite{Grozdanov:2017ajz, Blake:2017ris, Blake:2018leo}
\begin{align}
\label{eq:gravpsp}
\omega_* = +i \lambda_{L}, \qquad k_* = i \frac{\lambda_{L}}{v_B},
\end{align}
where $\lambda_{L} = 2 \pi T$ is the Lyapunov exponent for holographic theories and $v_B$ is the butterfly velocity.
Thus this location seems to be connected to the chaotic properties of holographic systems \cite{Larkin, Polchinski:2015cea,Maldacena:2015waa}%
\footnote{Pole-skipping in theories which are not maximally chaotic has been discussed in \cite{Ramirez:2020qer} and \cite{Choi:2020tdj}. In such theories the relation to  chaos is not straightforward and the applicability of pole-skipping as a diagnostic for chaos might be limited to only imposing bounds on chaos.},
which is the case even in the presence of stringy corrections \cite{Grozdanov:2018kkt} (see \cite{Natsuume:2019sfp,Abbasi:2019rhy, Liu:2020yaf,Abbasi:2020ykq, Jansen:2020hfd} for further examples).

Correlators associated to lower-spin fields also exhibit pole-skipping, albeit at different imaginary frequencies.
Scalar and $U(1)$ gauge fields were analysed in  \cite{Blake:2019otz} (see also \cite{Grozdanov:2019uhi,  Natsuume:2019xcy,Natsuume:2019vcv, Wu:2019esr,  Ahn:2020bks, Natsuume:2020snz}).
In the case of a scalar field  all pole-skipping points are located in the lower-half of the complex frequency plane at (imaginary) bosonic Matsubara frequencies  $\omega^B_n = - 2\pi i T n$,  where $n = 1,2, \ldots$.
The same applies for the gauge field, however in the longitudinal channel there exists an additional `hydrodynamic'  pole-skipping point located at $\omega = 0$ and $k =0$.
Note that such an infinite tower of pole-skipping points at negative imaginary Matsubara frequencies can also be found in the energy density correlator \cite{Blake:2019otz}.
Similarly, for fermion fields all pole-skipping points are on the lower-half of the complex frequency plane \cite{Ceplak:2019ymw}, but they are located at fermionic Matsubara frequencies $\omega^F_{n} =  - 2 \pi i T (n -1/2)$ where $n = 1,2, \ldots$.

Note that there is a relationship between the spins of bulk fields and the frequency values of the leading pole-skipping points as depicted in FIG.~\ref{figorder}.
The locations on the complex $\omega$-plane are given by $\omega_{\rm{0},s} = 2 \pi i T(s-1)$, where $s$ denotes the spin of the field%
\footnote{We thank  Richard Davison for bringing this order of leading pole-skipping points to our attention.
This kind of relation between the frequency and spin has also appeared in the analysis of
\cite{Perlmutter:2016pkf}, where holographic correlators (and correlation functions in theories with added higher spin currents) were studied.
Similarly, in \cite{Ahn:2020bks} (see also \cite{Kim:2020url}) the exchange of vector and scalar fields in a four-point correlation function on a hyperbolic space was considered and the leading Regge behaviour contained the same relation between the spin and the (imaginary) exponential coefficient multiplying the time coordinate.}. We note that FIG.~\ref{figorder} is a compilation of several independent results.
However, the analysis for spin-${3\ov 2}$ fields has been missing from the literature.
In this note, by analysing the Rarita-Schwinger field in the near-horizon region of an AdS-Schwarzschild black hole, we show that indeed the relation between the spin and value of the frequency of the leading pole-skipping point holds for any (half)-integer spin with $s\leq 2$.


In addition to the locations of the pole-skipping points, the near-horizon analysis also predicts the form of the Green's function near such special points.
Typically, the correlator near a generic pole-skipping location $(\omega_{\rm{ps}}, k_{\rm{ps}})$  takes the form \cite{Blake:2018leo, Blake:2019otz, Ceplak:2019ymw, Ahn:2020baf}
\begin{align}
\label{eq:psform}
G^R(\omega_{\rm{ps}} + \delta\omega, k_{\rm{ps}} + \delta k) \propto   \frac{\delta \omega - \left(\frac{\delta \omega}{\delta k}\right)_z \delta k }{\delta \omega - \left(\frac{\delta \omega}{\delta k}\right)_p \delta k }\,,
\end{align}
where $(\delta \omega/\delta k)_{p,z}$ denote the slope of the line of \emph{p}oles and line of \emph{z}eros passing through the pole-skipping point%
\footnote{A more thorough analysis of \cite{Ahn:2020baf} showed that there exist other forms that a Green's function can have near the special locations.
In their language, pole-skipping points at which the correlator takes on the form \eqref{eq:psform} are called \emph{Type-I}.
\emph{Type-II} pole-skipping points are associated with points that were previously denoted as anomalous points and are most commonly associated with locations at which the near-horizon analysis predicts two coincident pole-skipping points.
\emph{Type-III} pole-skipping points are associated with non-(half)-integer imaginary Matsubara frequencies and cannot be predicted from the near-horizon analysis (see also \cite{Das:2019tga}).}.
The Green's function of all leading pole-skipping points depicted in FIG.~\ref{figorder} has the form \eqref{eq:psform} and similarly we show that the near-horizon analysis predicts that the corresponding correlator at the leading pole-skipping point of the Rarita-Schwinger field  takes on the same form.

\section{Gravitational Setup}
\label{sec:rsc}


Let the bulk theory  be described by the Einstein-Hilbert action with a negative cosmological constant
\begin{align}\label{eq:backgroundaction}
S = \int d^{d+2}x \sqrt{- g} \left( R - 2 \Lambda \right)\,,
\end{align}
where $\Lambda = - d(d+1)/2L^2$ and $L$  denotes  the radius of AdS which we set to $L = 1$ in all further expressions.
The resulting equations of motion admit a planar  black hole solution that is asymptotically AdS and is described by the line element
\begin{align}\label{eq:backgroundmetricief}
ds^2 = - r^2 f(r) dv^2 + 2 dv\, dr +  h(r) dx^i\, dx^i\,.
\end{align}
We are using the {\ief} where  $r$ denotes the radial direction with the boundary of AdS located at $r \rightarrow \infty$.
The usual time coordinate can be recovered using the relation $v = t + r_*$, where $ dr_* =  dr/(r^2 f(r))$, in which case one can see that  $\{t,x^i\}$, with ${i=1\ldots d}$, are the coordinates of the $d+1$ dimensional Minkowski space at a fixed value of $r$.

The two functions appearing in the metric are given by
\begin{align}
\label{eq:ssfun}
f(r) = 1 - \left(\frac{r_0}{r}\right)^{d+1}\,, \qquad h(r)=r^2\,,
\end{align}
meaning that there is a non-degenerate horizon at a finite radius  $r= r_0$ where $f(r)$ vanishes.
However as we are using the {\ief}, such a point is a regular point of the metric.
The associated (non-vanishing) Hawking temperature is given by
$4 \pi T = r_0^2 f'(r_0) = (d+1)r_0 $.
In what follows, we keep  both $f(r)$ and $h(r)$ generic which allows us to  identify the source of the contributions  when analysing the location of pole-skipping points.
Furthermore, it allows for an easier generalisation of our results to a larger class of background geometries, such as geometries with an additional matter content deforming the background \cite{Andrade:2013gsa, Davison:2014lua}, even though our derivation applies only when $f(r)$ and $h(r)$ take the form \eqref{eq:ssfun}.


The aim of this note is the analysis of the  near-horizon behaviour of a minimally coupled spin-${3\ov 2}$ field. In an AdS/CFT context, such fields were first considered in \cite{Volovich:1998tj,Corley:1998qg,Koshelev:1998tu,Rashkov:1999ji,Matlock:1999fy} and have later been used to study the properties of the charged current in the dual boundary theory (see for example \cite{Policastro:2008cx, Gauntlett:2011mf, Gauntlett:2011wm, Erdmenger:2013thg}). For a summary and discussion of recent results see \cite{Liu:2013fja}.

The action describing the massive Rarita-Schwinger field $\Psi_M$  is given by%
\footnote{Throughout this note we use upper case Latin letters ($M, N, \ldots$) to denote the curved spacetime indices, whereas lower case Latin letters ($a,b, \ldots$) to denote the flat space indices.}
\begin{align}
\label{eq:RSac}
S_{RS} \propto \int d^{d+2}x \sqrt{-g}\, \bar{\Psi}_{M} \left( \Gamma^{MNP} \nabla_N - m\,\Gamma^{MP}\right) \Psi_P\,.
\end{align}
Since we are only interested in the bulk equations of motion, we do not need the precise details of the overall normalisation factor or the additional boundary terms%
\footnote{We note that another mass term  $m'g^{MN}\bar{\Psi}_M\,  \Psi_N$ can  be added to the action. This introduces a spin-{\half} degree of freedom in the Rarita-Schwinger field (see e.g. \cite{Liu:2013fja}) which may change the structure of pole-skipping. We do not consider such a term here.}.
 The anti-symmetrised products of curved space gamma matrices $\Gamma^{MP}$ and $\Gamma^{MNP}$ act on the spinor index of the Rarita-Schwinger field, which we suppress throughout the note.
The covariant derivative acting on the spin-${3\ov 2}$ field is given by
\mbox{$\nabla_M \, \Psi_P = \partial_M \, \Psi_P - \widetilde \Gamma^{N}_{MP}\,\Psi_N+ \frac{1}{4}\left(\omega_{ab}\right)_{M} \Gamma^{ab}\, \Psi_P$},
where $\widetilde  \Gamma^{N}_{MP}$ are the Christoffel symbols and  $\omega_M$ is the spin connection one form.

The equation of motion derived from \eqref{eq:RSac} is
\begin{align}
\label{eq:rseq1}
 \Gamma^{MNP}\, \nabla_N \,\Psi_P - m\, \Gamma^{MN}\, \Psi_N =0\,,
\end{align}
however since the background metric satisfies the vacuum Einstein's equation one can show (see for example \cite{Gauntlett:2011wm}) that the above equation of motion is equivalent to
\begin{align}
\label{eq:rseq2}
\left(\slashed \nabla + m\right) \Psi_N = 0\,,
\end{align}
with additional constraints
\begin{align}
\label{eq:cond1}
\Gamma^M\, \Psi_M =0\,, \qquad \nabla^M \, \Psi_M  = 0\,,
\end{align}
where we defined $\slashed \nabla = \Gamma^{M} \, \nabla_M$.
Note that while  \eqref{eq:rseq2} looks like a Dirac equation, it actually couples different vector components of the field due to the term involving the Christoffel symbols in the covariant derivative, which is not present if the covariant derivative is acting on a Dirac field.
In deriving \eqref{eq:rseq2} we assume that the mass takes on a generic  positive value in which case the constraints \eqref{eq:cond1} follow naturally from the equations of motion \eqref{eq:rseq1}.
If $m = d/2$, the spin-${3\ov 2}$ field is physically massless \cite{Deser:1977uq} (the non-zero value of $m$ is a consequence of the curvature of spacetime) and  the conditions \eqref{eq:cond1} are not imposed by the equations of motion, but can be considered as a choice of gauge.

\section{Pole-Skipping}
\label{sec:ps}

We now show that at a specific complex value of the frequency and momentum the equations of motion \eqref{eq:rseq2} and the associated constraints admit two independent solutions that are  regular at the horizon.
We use the following orthonormal frame%
\footnote{We use underlined indices to indicate specific flat space indices ($a=\uuv, \uur, \ldots$) and distinguish them from particular curved space indices which are not underlined ($M = v,r,\ldots$).}
\begin{gather}
E^{\uuv} =  \frac{1+ f(r)}{2}\,r dv - \frac{dr}{r}\,,\qquad E^{\uur} =  \frac{1- f(r)}{2}\, r dv + \frac{dr}{r}\,, \nonumber\\
 E^{\uui} = \sqrt{h(r)} \, dx^{i}\,,
 \label{eq:general_frame}
\end{gather}
in which case the line element \eqref{eq:backgroundmetricief} can be written as \mbox{$ds^2 = \eta_{ab} E^a\, E^b$} with \mbox{$\eta_{ab} = \text{diag}(-1, 1,1, \ldots,1)$}.

Next we use the fact that the  metric components depend only on the $r$ coordinate and introduce a plane wave ansatz $\Psi_M(v,r, x^i) = \psi_M(r)\,e^{-i \omega v + i k_i x^i}$ in which case the equations of motion reduce to a  system of coupled first order ordinary differential equations for $\psi_M(r)$.
To separate these equations into tractable subsystems we decompose the components of the field based on its eigenvalues under the action of  $\grr$ and $\g2 \equiv \hat k_i \gvi$, where $\hat k_i = k_i/k$ is the normalised Euclidean vector in $d$-dimensional flat space  and $k\equiv  \sqrt{\sum_{i=1}^d k_i^2}$ its magnitude.
These two matrices commute  and their eigenvalues are equal to $\pm1$, hence all vector components of the Rarita-Schwinger field can be decomposed as
\begin{align}
\label{eq:spdec}
\psi_M &= \sum_{\alpha_1 = \pm} \sum_{\alpha_2 = \pm} \psi_M^{(\alpha_1, \alpha_2)} 
\end{align}
where the indices in the bracket denote the eigenvalues under the action of the gamma matrices as
\begin{align}
	\label{eq:spdec2}
	\grr\, \psi_M^{(\alpha_1, \alpha_2)} = \alpha_1\, \psi_M^{(\alpha_1, \alpha_2)}, \quad \g2\, \psi_M^{( \alpha_1, \alpha_2)} = \alpha_2\, \psi_M^{( \alpha_1, \alpha_2)},
\end{align}
with $\alpha_{1,2} = \pm$.
Each of the components in the decomposition contains a quarter of the total degrees of freedom of the spinor.

The leading pole-skipping point for the energy-density correlator  is found by considering the $vv$-component of the dual bulk excitation \cite{Grozdanov:2017ajz, Blake:2017ris, Blake:2018leo}.
Similarly, the location of the leading pole-skipping point for the $U(1)$ gauge field is uncovered by considering the $v$-component of the bulk field \cite{Blake:2019otz}.
Expecting similar behaviour, we focus on the $\psi_v$ component of the Rarita-Schwinger field.
By using the constraints \eqref{eq:cond1}, we find that only $\psi_r$ couples to $\psi_v$ so we focus on that subsystem of equations.
This subsystem then separates into two decoupled sets of differential equation -- one containing components whose $\grr$ and $\g2$ eigenvalues are equal, and one with components whose eigenvalues under the action of the aforementioned matrices are opposite.
But the two decoupled systems of differential equations are related by  $k \rightarrow -k$. We can thus focus only on one and obtain the results for the other by a reversal of the momentum $k$.
For concreteness, we choose the subsystem dealing with the components whose eigenvalues under
$\grr$ and $\g2$ are equal.
These equations are given in full detail in the appendix (see \eqref{eq:eom4}), together with  some additional information about their derivation.

In our analysis the detailed form of the equations of motion is not needed as we are merely interested in their near-horizon expansion.
Because in the {\ief} the horizon is a regular point, we can expand the field components in a series as
\begin{align}
\label{eq:serexp}
\psi^{(\alpha_1, \alpha_2)}_M(r)= \sum_{l=0}^{\infty}\, \psi_{M,l}^{(\alpha_1, \alpha_2)}\, (r-r_0)^l\,,
\end{align}
where $\psi_{M,l}^{(\alpha_1, \alpha_2)}$ are constant coefficients.
Similarly, the equations of motion themselves can be expanded in a series at the horizon after which solving the differential equations translates into solving a system of algebraic equations at each order of the series expansion.

At a generic point in Fourier space these algebraic equations halve the number of free parameters in the Rarita-Schwinger field which corresponds to choosing the solution to the equations of motion that is ingoing at the horizon.
For example, by evaluating the equations of motion directly at the horizon, one finds, amongst others, the following equation (see \eqref{eq:seeom0})
\begin{align}
\label{eq:seeom0x}
  \nonumber
  &\left[ - 2 m r_0 - 4 i \omega + \frac{2 i k r_0 }{\sqrt{h(r_0)}}- r_0^2 f'(r_0)\right] \gvv \, \psi_{v,0}^{(-,-)} \\*
 &+  \left[ 2 m r_0 - 4 i \omega - \frac{2 i k r_0 }{\sqrt{h(r_0)}}- r_0^2 f'(r_0)\right] \psi_{v,0}^{(+,+)} = 0 \,,
\end{align}
which for generic $\omega$ and $k$ relates $\psi_{v,0}^{(-,-)}$ to $\psi_{v,0}^{(+,+)}$.
Similarly, equations of motion at higher order in the near-horizon expansion relate all other coefficients  (including those involving $\psi_r$ components) to the ones appearing in \eqref{eq:seeom0x} allowing us to perturbatively construct a solution with half of the total number of degrees of freedom.

Equation \eqref{eq:seeom0x} is trivially satisfied if both coefficients in the square brackets vanish, which is the case if the frequency and momentum are precisely
\begin{subequations}
\label{eq:psp1}
\begin{align}
\omega &\equiv \omega_0 = \frac{ir_0^2 f'(r_0)}{4}= i \pi T\,,\label{eq:spom1}\\*
  k &\equiv k_0 = - i m \sqrt{h(r_0)}= - \frac{4 m }{d+1} i  \pi T\label{eq:spk1}\,.
\end{align}
\end{subequations}
Then both $\psi_{v,0}^{(+,+)}$ and $\psi_{v,0}^{(-,-)}$ remain unconstrained and by using other algebraic equations from the near horizon expansion of the equations of motion  we can construct two linearly independent solutions that are regular at the horizon%
\footnote{If we set $\omega = i \pi T$ but leave $k$ generic, the complete near-horizon expansions, such as \eqref{eq:serexp}, contain additional logarithmic terms $~\log(r-r_0)$ similar to the terms appearing in the scalar \cite{Blake:2019otz} or fermion  \cite{Ceplak:2019ymw} field expansions.
However, one finds that by imposing the equations of motion,  all coefficients multiplying terms with a logarithmic divergence at the horizon have to vanish, unless the momentum is finely tuned to the pole-skipping value \eqref{eq:spk1}.
}.

\begin{figure}[th!]
\label{fig}
\centering
\includegraphics[width =0.77\columnwidth ]{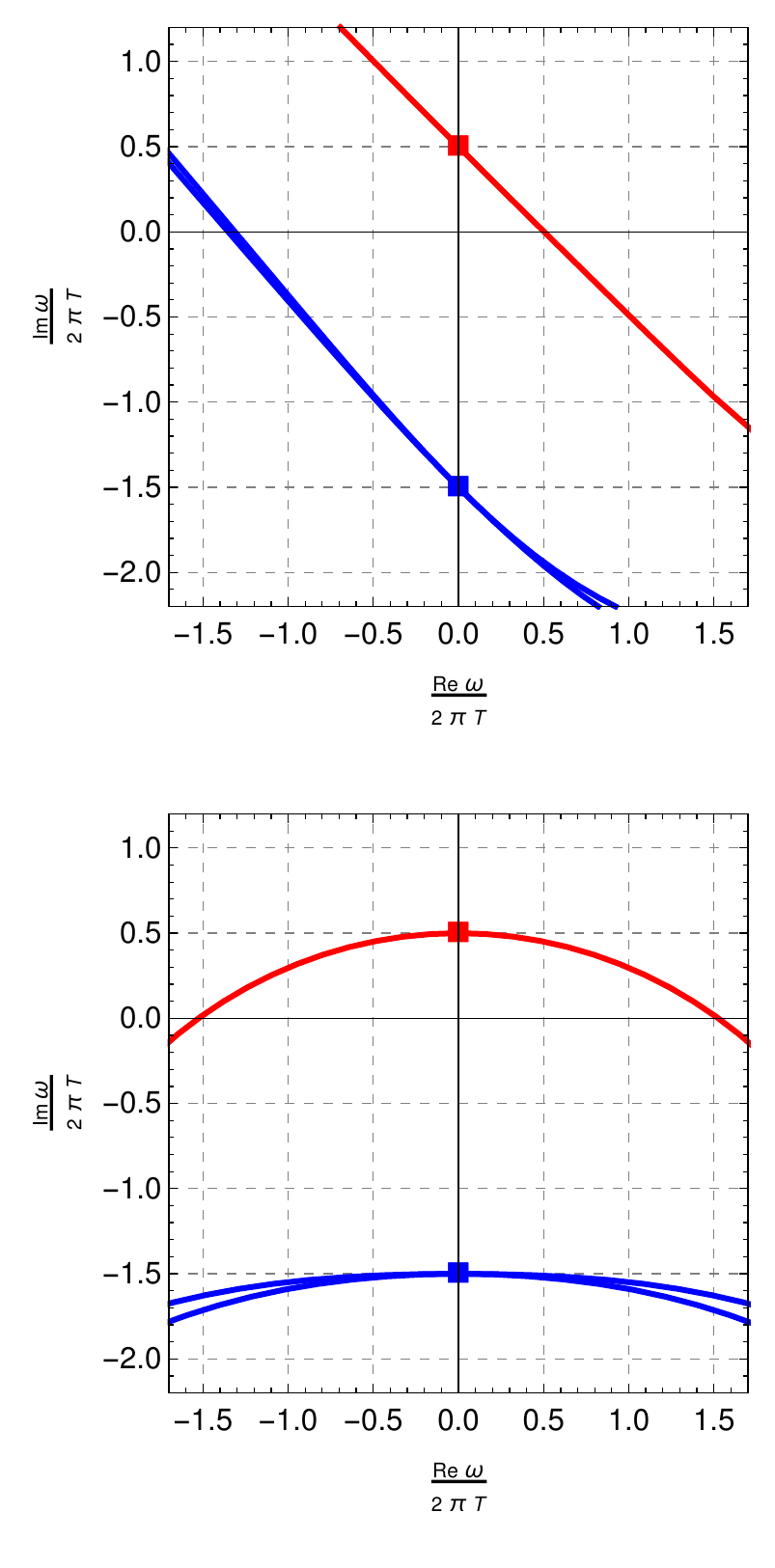}
\caption{Numerical analysis of the poles of the retarded Green's function in AdS$_4$ for a Rarita-Schwinger field with mass $m = 3.4$.
The poles are calculated using the methods presented in \cite{Denef:2009yy} adapted to the spin-$3/2$ case.
The top figure depicts the motion of the poles on the complex $\omega$-plane near the value of the momentum where pole-skipping is observed. The momentum is varied linearly as $k = k_0 + \epsilon e^{\frac{i \pi}{4}}$, where $\epsilon \in \mathbb{R}$.
We notice that the red curve passes through $\omega_0$ as $\epsilon =0$ which we denote with a red square.
In blue we depict  two additional poles which pass through $\omega = - 3 \pi i$ as $\epsilon \rightarrow 0$. These correspond to higher order pole-skipping points which are discussed in the appendix.
The bottom figure shows the location of the poles as the momentum is varied as  $k = k_0\,e^{ i\epsilon}$. Pole-skipping occurs at the filled squares.
}
\label{fig:graphs}
\end{figure}

The plots in FIG.~\ref{fig:graphs} show  the location of poles of the boundary Green's function in the complex $\omega$-plane at different values of the momentum, calculated using the Leaver method \cite{Denef:2009yy}. 
We see that at the special point \eqref{eq:spk1} there is a pole located exactly at \eqref{eq:spom1} -- this is the pole predicted by the near-horizon analysis.

Finally, recall that there exist an additional pole-skipping point at the same frequency \eqref{eq:spom1} but at the opposite momentum to \eqref{eq:spk1} that is obtained from the near-horizon analysis of the components, whose $\grr$ and $\g2$ eigenvalues are opposite.

As is known from previous results \cite{Blake:2017ris, Blake:2018leo, Blake:2019otz, Ceplak:2019ymw,Ahn:2020bks, Ahn:2020baf}, a location in Fourier space where there exist multiple independent ingoing solutions at the horizon corresponds  to a point where a pole and a zero of the boundary retarded Green's function coincide.
As such, \eqref{eq:psp1} is the first  location at which we observe pole-skipping for the Rarita-Schwinger field.
Most notably, this point is located at a \emph{positive} imaginary frequency.
Up to now, the only other example of a pole-skipping point being found on the upper complex $\omega$ half-plane is the leading pole-skipping location for the energy-density correlator, in which case the precise location in Fourier space has been conjectured to be related to the chaotic properties of the theory, as seen in \eqref{eq:gravpsp}.

For the Rarita-Schwinger field, the modulus \eqref{eq:spom1} is exactly half the value of the maximal Lyapunov exponent bound of \cite{Maldacena:2015waa}.
In our opinion this suggests that there is little relation to chaos, despite the seeming exponential growth of the solution, if the values  \eqref{eq:psp1} are inserted into the plane-wave ansatz.
This is further substantiated by the fact that the value of the momentum at which pole-skipping occurs depends on the mass of the field and thus the location is sensitive both to the background \emph{and} the probe.%

Finally, recall that the equations of motion near the horizon also predict the form of the Green's function near the pole-skipping point.
To that end evaluate \eqref{eq:seeom0x} at a nearby point in Fourier space, which we denote by $\omega = \omega_0 + \epsilon \delta\omega$ and $k = k_0 + \epsilon \delta k$, where $\epsilon >0$ is a small number used as a book-keeping device and we keep the direction of the momentum $\hat k$ fixed.
The analysis of \cite{Ceplak:2019ymw} applies to this case and we refer the reader to that reference for a detailed calculation.
The final result of \cite{Ceplak:2019ymw} applies here as well: working at linear order in $\epsilon$, one finds that near \eqref{eq:psp1} the retarded Green's function takes the pole-skipping form \eqref{eq:psform}.

\section{Discussion}
\label{sec:disc}

It has recently been shown that quantum chaos in field theories with holographic duals manifests itself in the thermal energy density two-point functions. These correlators exhibit the `pole-skipping' phenomenon at special (imaginary) values of the frequency and momentum. At such points poles and zeroes collide and the Green's function becomes ill-defined. It was found \cite{Grozdanov:2017ajz, Blake:2017ris, Blake:2018leo} that the pole-skipping frequency lies at $\omega = + 2 \pi i T$. The modulus $|\om|=2\pi T$ is thought to be related to the maximal Lyapunov exponent \cite{Maldacena:2015waa}.

\comment{
\begin{figure}[h!]
\centering
\includegraphics[width =0.4\columnwidth ]{fig02.jpg}
\caption{\label{figcap} The Euclidean black hole geometry caps off smoothly in the interior. In the figure, a spin-2 field configuration is indicated as a deformation of the geometry.
A shift along the Euclidean time direction (arrow) translates to a rotation at the tip.
 }
\end{figure}
}

For bulk fields with spins other than two, the corresponding boundary two-point functions also show pole-skipping at certain Matsubara frequencies. Scalar fields first exhibit pole-skipping at $\omega = - 2 \pi i T$, while
for conserved currents this occurs at $\omega = 0$ \cite{Blake:2019otz}. Fermionic fields show pole-skipping at $\omega = - \pi i T$, i.e. at the first negative fermionic Matsubara frequency \cite{Ceplak:2019ymw}.
The calculations rely on an analysis of the near-horizon region of the bulk geometry and give non-trivial constraints on Green's functions at frequencies $\om \sim T$.

Although the relevant pole-skipping momenta do not seem to be universal, there is a definite structure in  pole-skipping frequencies; see FIG. \ref{figorder}. Namely, there is a relationship between the frequency of the first pole-skipping point and the spin of the bulk field.

In this paper we have dealt with the spin-$3\ov 2$ case which has been missing from previous analyses.
In particular, we have shown from a bulk perspective that a Rarita-Schwinger field in an AdS-Schwarzschild background exhibits pole-skipping precisely at the expected first fermionic Matsubara frequency on the upper-half plane.
Since the analysis only concerned the near-horizon region, we expect that the results can be extended to more general spacetimes with regular horizons.
With this result, we complete the hierarchy of pole-skipping locations for various fields of different spin.

It is important to note that the results for individual fields were calculated independently and in no way relied on supersymmetry.
Furthermore, the above mentioned locations are merely the first pole-skipping points  and in some sense are the simplest ones - the locations in momentum space depend only on the values of the background metric at the horizon and/or the nearby region through first derivatives, whereas higher order pole-skipping points generically depend on higher order derivatives as well \cite{Blake:2019otz}.
Thus it is enticing to conjecture that the hierarchy is the property of the near-horizon region of spacetime itself.

It would be interesting to investigate the origin of this hierarchy (for a CFT analysis, see  \cite{Haehl:2018izb, Das:2019tga, Haehl:2019eae}).
Note that the pole-skipping points are located at (imaginary) Matsubara frequencies which means that the static bulk has a Euclidean counterpart (obtained by Wick-rotation). 
Since the geometry smoothly caps off, a shift along the Euclidean time circle translates into a rotation at the tip (which is the Euclidean analog of the event horizon).  Fourier modes are therefore connected to spin, providing an explanation for the observed hierarchy of pole-skipping points.

In the above analysis we have merely shown the existence of a single pole-skipping point, but previous results on other fields would suggest that there exists an entire tower of higher-order pole-skipping points at negative imaginary frequencies, which we have not focused on.
Furthermore, the above analysis can be expanded to include matter fields or allow for an additional mass term in the action of the Rarita-Schwinger field. It would be interesting to see if there exist a configuration at which the pole-skipping points vanish.

Finally, the Rarita-Schwinger fields in backgrounds with simple horizons have been used in calculations of fermionic currents  \cite{Policastro:2008cx, Gauntlett:2011mf, Gauntlett:2011wm, Erdmenger:2013thg}. It would be interesting to study the pole-skipping points in that context and see the consequences on the spectral function of the dual fermionic currents in the boundary theory.

\vspace{0.2in}   \centerline{\bf{ACKNOWLEDGEMENTS}} \vspace{0.2in}

We are grateful to  Richard Davison for helpful discussions.
N\v{C} is supported by the ERC Grant 787320 - QBH Structure.
DV is supported by the STFC Ernest Rutherford grant ST/P004334/1.

\appendix

\section{Equations of motion}
\label{app:conv}
In this appendix we spell out some of the details that have been omitted in the main part of this note.
We keep the same notational conventions as in the main text.
In addition, throughout the note we use the following convention for the flat space Clifford algebra
\begin{align}
\left\{ \Gamma^a, \Gamma^b\right\} = 2 \eta^{ab}\,,
\end{align}
where $\eta^{ab} = \text{diag}(-1, +1,  \ldots, +1)$, which most importantly implies that $\gvv$ squares to $-1$ while all others square to $1$.

We start at the rewritten equations of motion \eqref{eq:rseq2}.
We again want to point out that we keep $h(r)$ and $f(r)$ unevaluated to keep track of the origin of individual terms in the final expressions. However, in deriving \eqref{eq:rseq2} from \eqref{eq:rseq1}, one assumes that the Ricci tensor and the Ricci scalar satisfy
\begin{align}\label{eq:rcond}
R_{MN} = -(d+1)g_{MN}\,, \qquad R = -(d+1)(d+2)\,,
\end{align}
which is the case if the two aforementioned functions take the form \eqref{eq:ssfun}, as for example for a black brane solution.
So while the results presented in this note hold only for backgrounds that satisfy \eqref{eq:rcond}, we believe that the generalisation of our results to more complicated backgrounds should be straightforward and that the findings would resemble those of this note.

Recall that due to the presence of the Christoffel terms in the covariant derivative, different vector components of the Rarita-Schwinger field are coupled.
We can make this explicit by putting such terms on the right-hand side, resulting in
\begin{align}
\label{eq:rseq3}
\left(\slashed D + m\right) \Psi_M = \Gamma^{N}\, \widetilde \Gamma^{P}_{MN} \, \Psi_P\,,
\end{align}
where $D_M = \partial_M +  \frac{1}{4}\left(\omega_{ab}\right)_{M} \Gamma^{ab}$ denotes the covariant derivative acting on a spinor field.
The left-hand side of equation \eqref{eq:rseq3} with the choice of vielbein \eqref{eq:general_frame} has been worked out in \cite{Ceplak:2019ymw}.
One can then use  the gamma-traceless condition \eqref{eq:cond1} to arrive at two coupled equations containing only the components $\Psi_v$ and $\Psi_r$
\begin{subequations}
\label{eq:eom2}
\begin{align}
&\left(\slashed D + m \right) \Psi_v = \frac{\partial_r (r^2 f(r))}{2r}\biggr\{\bigr[\gvv+ \grr\bigr]\Psi_v \nonumber\\*
& \qquad\quad \left.+ \frac{r^2}{2}\bigr[ (1+f(r))\gvv-\left(1-f(r)\right) \grr\bigr]\Psi_r \right\}\,, \label{eq:eom2v}\\\nonumber \\
&\left(\slashed D + m \right) \Psi_r =-\frac{\partial_r (r^2 f(r))}{2r} \bigr[\gvv+ \grr\bigr]\Psi_r - \frac{\partial_r h(r)}{2 r h(r)}\times \nonumber\\*
&\left\{ \bigr[\gvv+ \grr\bigr] \Psi_v + \frac{r^2}{2}\biggr[ (1+ f(r)) \grr-(1-f(r)) \gvv \biggr] \Psi_r\right\}
.\label{eq:eom2r}
\end{align}
\end{subequations}
Since the components of the metric depend only on the coordinate $r$, one can write the field in the form of a plane wave $\Psi_M(r) = \psi_M(r)\,e^{-i \omega v + i k_i x^i}$, and insert this ansatz into the above equations.
After decomposing the spinors in terms of their behaviour under the action of $\grr$ and $\g2\equiv \hat k_i \gvi $ matrices, as described in \eqref{eq:spdec} and \eqref{eq:spdec2}, one obtains eight coupled first order ordinary differential equations for the different components of $\psi_v$ and $\psi_r$.
The system of equations can be split into two decoupled subsystems of 4 equations, with one describing the components whose $\grr$ and $\g2$ eigenvalues are equal and the other containing the components with opposite eigenvalues under the action of the two matrices.
The two subsystems of equations are related by $k\rightarrow -k$, so it is sufficient to analyse only one of the subsystems and infer the results of the other by simply reversing the momentum.

Hence in what follows we focus at the system of equations that involve $\psi_v^{(\pm, \pm)}$ and $\psi_{r}^{(\pm, \pm)}$ which are given, after some algebra, by equations
%
\begin{widetext}
\begin{subequations}
\label{eq:eom4}
\begin{align}
\cS_1 &= \left[r^2 f(r) \partial_r - i \omega - \frac{2 r f(r) + r^2 f'(r) }{4}+ \frac{d\,  r^2 f(r) h'(r) }{4 h(r) }+ \frac{m r(1+ f(r))}{2}- \frac{i k r (1-f(r))}{2 \sqrt{h(r)}}\right] \vg{+}\nonumber \\*
&+ \gvv \left[ - i \omega - \frac{4 r f(r) + r^2 f'(r)}{4}- \frac{m r (1- f(r))}{2}+ \frac{i k r (1+ f(r))}{2 \sqrt{h(r)}} \right] \vg{-}-\frac{r^2 f(r) }{2}\partial_r(r^2 f(r))\gvv\, \rg{-}\,,\\
\cS_2&= \gvv\left[ r^2 f(r) \partial_r - i \omega - \frac{2r f(r)+ r^2 f'(r)}{4}+ \frac{d\, r^2 f(r) h'(r)}{4 h(r) }- \frac{m r (1+ f(r))}{2}+ \frac{i k r (1- f(r))}{2 \sqrt{h(r)}}\right] \vg{-}\nonumber\\*
&+\left[- i \omega - \frac{4 r f(r) + r^2 f'(r) }{4}+ \frac{m r(1- f(r))}{2}- \frac{i k r (1+ f(r))}{2 \sqrt{h(r)} } \right] \vg{+} - \frac{r^2 f(r) }{2}\partial_r(r^2 f(r))\, \rg{+}\,,\\
\cS_3 &= \left[r^2 f(r) \partial_r - i \omega + \frac{3(2 r f(r) + r^2 f'(r)) }{4}+ \frac{(d+2)\,  r^2 f(r) h'(r) }{4 h(r) }+ \frac{m r(1+ f(r))}{2}- \frac{i k r (1-f(r))}{2 \sqrt{h(r)}} \right] \rg{+}\nonumber \\
&+ \gvv \left[ - i \omega + \frac{4 r f(r) +3 r^2 f'(r)}{4}- \frac{m r (1- f(r))}{2}+ \frac{i k r (1+ f(r))}{2 \sqrt{h(r)}} \right] \rg{-}+\frac{h'(r)}{2h(r) }\left( \vg{+}+ \gvv \, \vg{-}\right) \,,\\
\cS_4 &=  \gvv \left[r^2 f(r) \partial_r - i \omega + \frac{3(2 r f(r) + r^2 f'(r)) }{4}+ \frac{(d+2)\,  r^2 f(r) h'(r) }{4 h(r) }- \frac{m r(1+ f(r))}{2}+ \frac{i k r (1-f(r))}{2 \sqrt{h(r)}} \right]\rg{-}\nonumber\\*
&+ \left[ - i \omega + \frac{4 r f(r) +3 r^2 f'(r)}{4}+ \frac{m r (1- f(r))}{2}- \frac{i k r (1+ f(r))}{2 \sqrt{h(r)}} \right] \rg{+}+ \frac{h'(r)}{2h(r) }\left( \vg{+}+ \gvv \, \vg{-}\right) \,.
\end{align}
\end{subequations}
\end{widetext}
For later convenience we have labelled them as $\cS_i$, with $i =1,2,3,4$, so that the equations of motion can be summarized in a compact way as $\cS_i =0$.

We are interested in the near-horizon expansion of these equations.
As we are working with the {\ief} in which the horizon is a regular point, we assume that all functions and fields can be expanded in a series around the horizon at $r = r_0$.
The expansion of \eqref{eq:ssfun} is trivial and we use the field expansion \eqref{eq:serexp}.
Then \eqref{eq:eom4} also get expanded around the horizon
\begin{align}
\label{eq:sereom}
\cS_i&=\sum_{l=0}^{\infty}\cS_{i}^{(l)} \, (r-r_0)^l = 0,\;  \Rightarrow \;  \cS_i^{(l)} = 0\,, 
\quad   i =1,2,3,4,
\end{align}
in which case the equations of motion become an infinite set of algebraic equations that we can solve order by order.

Our main interest lies in  evaluating the equations of motion directly at the horizon, or in other words, the zeroth order equations $\cS_i^{(0)} = 0$.
While the four equations  in \eqref{eq:eom4} are independent in general, one finds that directly at the horizon there are only two independent equations as $\cS^{(0)}_1 = \cS_2^{(0)}$ and $\cS^{(0)}_3 = \cS^{(0)}_4$, with
\begin{subequations}
\label{eq:seeom0}
\begin{align}
\cS_1^{(0)} &=\left[ - 2 m r_0 - 4 i \omega + \frac{2 i k r_0 }{\sqrt{h(r_0)}}- r_0^2 f'(r_0)\right] \gvv \, \vz{-}
\nonumber\\*
&\;+ \left[ 2 m r_0 - 4 i \omega - \frac{2 i k r_0 }{\sqrt{h(r_0)}}- r_0^2 f'(r_0)\right] \vz{+} = 0,\label{eq:seeom0a} \\
\cS_3^{(0)} &=   \left[-2 m r_0 -4 i \omega + \frac{2 i k r_0 }{\sqrt{h(r_0)}}+ 3 r_0^2 f'(r_0)\right]\gvv\, \rz{-}
 \nonumber\\*
& \quad +\left[2 m r_0 - 4 i \omega - \frac{2 i k r_0 }{\sqrt{h(r_0)}}+ 3 r_0^2 f'(r_0)\right] \rz{+} \nonumber\\*
&\quad + \frac{2h'(r_0)}{ h(r_0)}\left(\vz{+}+\gvv \, \vz{-}\right) = 0\label{eq:seeom0b} \,.
\end{align}
\end{subequations}
In fact \eqref{eq:seeom0a} is presented in the main text as \eqref{eq:seeom0x}.

In order to find the locations of pole-skipping points, we need to look for the values of $\omega$ and $k$ at which \eqref{eq:seeom0} and their higher-order analogues do not impose enough constraints on the solutions of the equations of motion to uniquely determine the retarded Green's function of the boundary theory.
The determination of the leading pole-skipping point is described in the main text (see the discussion around equation \eqref{eq:psp1}), but in fact there also exists an infinite tower of pole-skipping points at \emph{negative} imaginary fermionic Matsubara frequencies  $\omega^F_{n} =  - 2 \pi i T (n -1/2)$ where $n = 1,2, \ldots$.

The procedure of identifying such locations mimics the analysis presented for the fermion case \cite{Ceplak:2019ymw}, but with some additional caveats, and wont be discussed in detail here.
We just report some partial results which explain the additional poles found in the numerical analysis that we present in FIG.~\ref{fig:graphs}.
In order to find the pole-skipping points at $\omega_1^F = - i \pi T$, one needs to analyse the equations at linear order in the expansion \eqref{eq:sereom} and the locations are obtained by looking for points at which a linear combination of $\psi_{v,1}^{(+,+)}$ and $\psi_{v,1}^{(-,-)}$ represents an additional free parameter of the solutions to the equations of motion.
We find that for a generic value of the mass $m$ there are three pole-skipping points at this value of the frequency.

Pole-skipping points at $\omega_2^F = - 3i \pi T$ are perhaps more interesting as they arise due to both  $\psi_{v,2}^{(\alpha_1, \alpha_2)}$ and $\psi_{r,0}^{(\alpha_1, \alpha_2)}$.
In fact, one finds that two special points coincide at
\begin{subequations}
\label{eq:psp2}
\begin{align}
\omega &= - \frac{3i}{4}r_0^2 f'(r_0)= - 3 i \pi T\,,\label{eq:spom2}\\
k &= - i m \sqrt{h(r_0)}= - \frac{4i  \pi }{d+1}m  T\,,\label{eq:spk2}
\end{align}
\end{subequations}
and are thus responsible for the additional two poles that can be seen in FIG.~\ref{fig:graphs}.
A simple way to confirm that this is indeed the case is to insert these values into \eqref{eq:seeom0} and observe that the prefactors in the square brackets  \eqref{eq:seeom0b} vanish, hence leaving $\rz{+}$ and $\rz{-}$ unconstrained. Furthermore, one finds that at these values of the frequency and momentum \eqref{eq:seeom0a} and \eqref{eq:seeom0b} give equivalent constraints.
We have not investigated the significance of this ``double'' pole-skipping point (see the analysis of \cite{Ahn:2020baf}), but we attribute this occurrence to the interplay between the $\psi_r$ and $\psi_v$ components in the equations of motion.

\bibliographystyle{utphys}

\bibliography{gps}

\providecommand{\href}[2]{#2}\begingroup\raggedright\begin{thebibliography}{10}

\bibitem{Maldacena:1997re}
J.~M. Maldacena, ``{The Large N limit of superconformal field theories and
  supergravity},'' \href{http://dx.doi.org/10.1023/A:1026654312961,
  10.4310/ATMP.1998.v2.n2.a1}{{\em Int. J. Theor. Phys.} {\bfseries 38} (1999)
  1113--1133}, \href{http://arxiv.org/abs/hep-th/9711200}{{\ttfamily
  arXiv:hep-th/9711200 [hep-th]}}.
[Adv. Theor. Math. Phys.2,231(1998)].

\bibitem{Gubser:1998bc}
S.~S. Gubser, I.~R. Klebanov, and A.~M. Polyakov, ``{Gauge theory correlators
  from noncritical string theory},''
  \href{http://dx.doi.org/10.1016/S0370-2693(98)00377-3}{{\em Phys. Lett.}
  {\bfseries B428} (1998) 105--114},
\href{http://arxiv.org/abs/hep-th/9802109}{{\ttfamily arXiv:hep-th/9802109
  [hep-th]}}.

\bibitem{Witten:1998qj}
E.~Witten, ``{Anti-de Sitter space and holography},''
  \href{http://dx.doi.org/10.4310/ATMP.1998.v2.n2.a2}{{\em Adv. Theor. Math.
  Phys.} {\bfseries 2} (1998) 253--291},
\href{http://arxiv.org/abs/hep-th/9802150}{{\ttfamily arXiv:hep-th/9802150
  [hep-th]}}.

\bibitem{Son:2002sd}
D.~T. Son and A.~O. Starinets, ``{Minkowski space correlators in AdS / CFT
  correspondence: Recipe and applications},''
  \href{http://dx.doi.org/10.1088/1126-6708/2002/09/042}{{\em JHEP} {\bfseries
  09} (2002) 042},
\href{http://arxiv.org/abs/hep-th/0205051}{{\ttfamily arXiv:hep-th/0205051
  [hep-th]}}.

\bibitem{Kovtun:2012rj}
P.~Kovtun, ``{Lectures on hydrodynamic fluctuations in relativistic
  theories},'' \href{http://dx.doi.org/10.1088/1751-8113/45/47/473001}{{\em J.
  Phys.} {\bfseries A45} (2012) 473001},
\href{http://arxiv.org/abs/1205.5040}{{\ttfamily arXiv:1205.5040 [hep-th]}}.

\bibitem{Glorioso:2018wxw}
H.~Liu and P.~Glorioso, ``{Lectures on non-equilibrium effective field theories
  and fluctuating hydrodynamics},''
  \href{http://dx.doi.org/10.22323/1.305.0008}{{\em PoS} {\bfseries TASI2017}
  (2018) 008}, \href{http://arxiv.org/abs/1805.09331}{{\ttfamily
  arXiv:1805.09331 [hep-th]}}.

\bibitem{Iqbal:2008by}
N.~Iqbal and H.~Liu, ``{Universality of the hydrodynamic limit in AdS/CFT and
  the membrane paradigm},''
  \href{http://dx.doi.org/10.1103/PhysRevD.79.025023}{{\em Phys. Rev.}
  {\bfseries D79} (2009) 025023},
\href{http://arxiv.org/abs/0809.3808}{{\ttfamily arXiv:0809.3808 [hep-th]}}.

\bibitem{Kovtun:2004de}
P.~Kovtun, D.~T. Son, and A.~O. Starinets, ``{Viscosity in strongly interacting
  quantum field theories from black hole physics},''
  \href{http://dx.doi.org/10.1103/PhysRevLett.94.111601}{{\em Phys. Rev. Lett.}
  {\bfseries 94} (2005) 111601},
\href{http://arxiv.org/abs/hep-th/0405231}{{\ttfamily arXiv:hep-th/0405231
  [hep-th]}}.

\bibitem{Grozdanov:2017ajz}
S.~Grozdanov, K.~Schalm, and V.~Scopelliti, ``{Black hole scrambling from
  hydrodynamics},''
  \href{http://dx.doi.org/10.1103/PhysRevLett.120.231601}{{\em Phys. Rev.
  Lett.} {\bfseries 120} no.~23, (2018) 231601},
\href{http://arxiv.org/abs/1710.00921}{{\ttfamily arXiv:1710.00921 [hep-th]}}.

\bibitem{Blake:2017ris}
M.~Blake, H.~Lee, and H.~Liu, ``{A quantum hydrodynamical description for
  scrambling and many-body chaos},''
  \href{http://dx.doi.org/10.1007/JHEP10(2018)127}{{\em JHEP} {\bfseries 10}
  (2018) 127},
\href{http://arxiv.org/abs/1801.00010}{{\ttfamily arXiv:1801.00010 [hep-th]}}.

\bibitem{Blake:2018leo}
M.~Blake, R.~A. Davison, S.~Grozdanov, and H.~Liu, ``{Many-body chaos and
  energy dynamics in holography},''
  \href{http://dx.doi.org/10.1007/JHEP10(2018)035}{{\em JHEP} {\bfseries 10}
  (2018) 035},
\href{http://arxiv.org/abs/1809.01169}{{\ttfamily arXiv:1809.01169 [hep-th]}}.

\bibitem{Larkin}
A.~I. Larkin and Y.~N. Ovchinnikov, ``{Quasiclassical method in the theory of
  superconductivity},'' {\em JETP} {\bfseries 28, 6} (1969) 1200--1205.

\bibitem{Polchinski:2015cea}
J.~Polchinski, ``{Chaos in the black hole S-matrix},''
\href{http://arxiv.org/abs/1505.08108}{{\ttfamily arXiv:1505.08108 [hep-th]}}.

\bibitem{Maldacena:2015waa}
J.~Maldacena, S.~H. Shenker, and D.~Stanford, ``{A bound on chaos},''
  \href{http://dx.doi.org/10.1007/JHEP08(2016)106}{{\em JHEP} {\bfseries 08}
  (2016) 106},
\href{http://arxiv.org/abs/1503.01409}{{\ttfamily arXiv:1503.01409 [hep-th]}}.

\bibitem{Ramirez:2020qer}
D.~M. Ramirez, ``{Chaos and pole skipping in CFT$_2$},''
  \href{http://arxiv.org/abs/2009.00500}{{\ttfamily arXiv:2009.00500
  [hep-th]}}.

\bibitem{Choi:2020tdj}
C.~Choi, M.~Mezei, and G.~S\'arosi, ``{Pole skipping away from maximal
  chaos},'' \href{http://arxiv.org/abs/2010.08558}{{\ttfamily arXiv:2010.08558
  [hep-th]}}.

\bibitem{Grozdanov:2018kkt}
S.~Grozdanov, ``{On the connection between hydrodynamics and quantum chaos in
  holographic theories with stringy corrections},''
  \href{http://dx.doi.org/10.1007/JHEP01(2019)048}{{\em JHEP} {\bfseries 01}
  (2019) 048},
\href{http://arxiv.org/abs/1811.09641}{{\ttfamily arXiv:1811.09641 [hep-th]}}.

\bibitem{Natsuume:2019sfp}
M.~Natsuume and T.~Okamura, ``{Holographic chaos, pole-skipping, and
  regularity},''
\href{http://arxiv.org/abs/1905.12014}{{\ttfamily arXiv:1905.12014 [hep-th]}}.

\bibitem{Abbasi:2019rhy}
N.~Abbasi and J.~Tabatabaei, ``{Quantum chaos, pole-skipping and hydrodynamics
  in a holographic system with chiral anomaly},''
\href{http://arxiv.org/abs/1910.13696}{{\ttfamily arXiv:1910.13696 [hep-th]}}.

\bibitem{Liu:2020yaf}
Y.~Liu and A.~Raju, ``{Quantum Chaos in Topologically Massive Gravity},''
  \href{http://arxiv.org/abs/2005.08508}{{\ttfamily arXiv:2005.08508
  [hep-th]}}.

\bibitem{Abbasi:2020ykq}
N.~Abbasi and S.~Tahery, ``{Complexified quasinormal modes and the
  pole-skipping in a holographic system at finite chemical potential},''
  \href{http://dx.doi.org/10.1007/JHEP10(2020)076}{{\em JHEP} {\bfseries 10}
  (2020) 076}, \href{http://arxiv.org/abs/2007.10024}{{\ttfamily
  arXiv:2007.10024 [hep-th]}}.

\bibitem{Jansen:2020hfd}
A.~Jansen and C.~Pantelidou, ``{Quasinormal modes in charged fluids at complex
  momentum},'' \href{http://dx.doi.org/10.1007/JHEP10(2020)121}{{\em JHEP}
  {\bfseries 10} (2020) 121}, \href{http://arxiv.org/abs/2007.14418}{{\ttfamily
  arXiv:2007.14418 [hep-th]}}.

\bibitem{Blake:2019otz}
M.~Blake, R.~A. Davison, and D.~Vegh, ``{Horizon constraints on holographic
  Green's functions},''
\href{http://arxiv.org/abs/1904.12883}{{\ttfamily arXiv:1904.12883 [hep-th]}}.

\bibitem{Grozdanov:2019uhi}
S.~s. Grozdanov, P.~K. Kovtun, A.~O. Starinets, and P.~Tadi\'c, ``{The complex
  life of hydrodynamic modes},''
  \href{http://dx.doi.org/10.1007/JHEP11(2019)097}{{\em JHEP} {\bfseries 11}
  (2019) 097}, \href{http://arxiv.org/abs/1904.12862}{{\ttfamily
  arXiv:1904.12862 [hep-th]}}.

\bibitem{Natsuume:2019xcy}
M.~Natsuume and T.~Okamura, ``{Nonuniqueness of Green's functions at special
  points},''
\href{http://arxiv.org/abs/1905.12015}{{\ttfamily arXiv:1905.12015 [hep-th]}}.

\bibitem{Natsuume:2019vcv}
M.~Natsuume and T.~Okamura, ``{Pole-skipping with finite-coupling
  corrections},'' \href{http://dx.doi.org/10.1103/PhysRevD.100.126012}{{\em
  Phys. Rev. D} {\bfseries 100} no.~12, (2019) 126012},
  \href{http://arxiv.org/abs/1909.09168}{{\ttfamily arXiv:1909.09168
  [hep-th]}}.

\bibitem{Wu:2019esr}
X.~Wu, ``{Higher curvature corrections to pole-skipping},''
  \href{http://dx.doi.org/10.1007/JHEP12(2019)140}{{\em JHEP} {\bfseries 12}
  (2019) 140}, \href{http://arxiv.org/abs/1909.10223}{{\ttfamily
  arXiv:1909.10223 [hep-th]}}.

\bibitem{Ahn:2020bks}
Y.~Ahn, V.~Jahnke, H.-S. Jeong, K.-Y. Kim, K.-S. Lee, and M.~Nishida,
  ``{Pole-skipping of scalar and vector fields in hyperbolic space: conformal
  blocks and holography},'' \href{http://arxiv.org/abs/2006.00974}{{\ttfamily
  arXiv:2006.00974 [hep-th]}}.

\bibitem{Natsuume:2020snz}
M.~Natsuume and T.~Okamura, ``{Pole-skipping and zero temperature},''
  \href{http://arxiv.org/abs/2011.10093}{{\ttfamily arXiv:2011.10093
  [hep-th]}}.

\bibitem{Ceplak:2019ymw}
N.~Ceplak, K.~Ramdial, and D.~Vegh, ``{Fermionic pole-skipping in
  holography},'' \href{http://dx.doi.org/10.1007/JHEP07(2020)203}{{\em JHEP}
  {\bfseries 07} (2020) 203}, \href{http://arxiv.org/abs/1910.02975}{{\ttfamily
  arXiv:1910.02975 [hep-th]}}.

\bibitem{Perlmutter:2016pkf}
E.~Perlmutter, ``{Bounding the Space of Holographic CFTs with Chaos},''
  \href{http://dx.doi.org/10.1007/JHEP10(2016)069}{{\em JHEP} {\bfseries 10}
  (2016) 069}, \href{http://arxiv.org/abs/1602.08272}{{\ttfamily
  arXiv:1602.08272 [hep-th]}}.

\bibitem{Kim:2020url}
K.-Y. Kim, K.-S. Lee, and M.~Nishida, ``{Holographic scalar and vector exchange
  in OTOCs and pole-skipping phenomena},''
  \href{http://arxiv.org/abs/2011.13716}{{\ttfamily arXiv:2011.13716
  [hep-th]}}.

\bibitem{Ahn:2020baf}
Y.~Ahn, V.~Jahnke, H.-S. Jeong, K.-Y. Kim, K.-S. Lee, and M.~Nishida,
  ``{Classifying pole-skipping points},''
  \href{http://arxiv.org/abs/2010.16166}{{\ttfamily arXiv:2010.16166
  [hep-th]}}.

\bibitem{Das:2019tga}
S.~Das, B.~Ezhuthachan, and A.~Kundu, ``{Real Time Dynamics in Low Point
  Correlators},''
\href{http://arxiv.org/abs/1907.08763}{{\ttfamily arXiv:1907.08763 [hep-th]}}.

\bibitem{Andrade:2013gsa}
T.~Andrade and B.~Withers, ``{A simple holographic model of momentum
  relaxation},'' \href{http://dx.doi.org/10.1007/JHEP05(2014)101}{{\em JHEP}
  {\bfseries 05} (2014) 101},
\href{http://arxiv.org/abs/1311.5157}{{\ttfamily arXiv:1311.5157 [hep-th]}}.

\bibitem{Davison:2014lua}
R.~A. Davison and B.~Gouteraux, ``{Momentum dissipation and effective theories
  of coherent and incoherent transport},''
  \href{http://dx.doi.org/10.1007/JHEP01(2015)039}{{\em JHEP} {\bfseries 01}
  (2015) 039},
\href{http://arxiv.org/abs/1411.1062}{{\ttfamily arXiv:1411.1062 [hep-th]}}.

\bibitem{Volovich:1998tj}
A.~Volovich, ``{Rarita-Schwinger field in the AdS / CFT correspondence},''
  \href{http://dx.doi.org/10.1088/1126-6708/1998/09/022}{{\em JHEP} {\bfseries
  09} (1998) 022},
\href{http://arxiv.org/abs/hep-th/9809009}{{\ttfamily arXiv:hep-th/9809009
  [hep-th]}}.

\bibitem{Corley:1998qg}
S.~Corley, ``{The Massless gravitino and the AdS / CFT correspondence},''
  \href{http://dx.doi.org/10.1103/PhysRevD.59.086003}{{\em Phys. Rev.}
  {\bfseries D59} (1999) 086003},
\href{http://arxiv.org/abs/hep-th/9808184}{{\ttfamily arXiv:hep-th/9808184
  [hep-th]}}.

\bibitem{Koshelev:1998tu}
A.~S. Koshelev and O.~A. Rytchkov, ``{Note on the massive Rarita-Schwinger
  field in the AdS / CFT correspondence},''
  \href{http://dx.doi.org/10.1016/S0370-2693(99)00148-3}{{\em Phys. Lett.}
  {\bfseries B450} (1999) 368--376},
\href{http://arxiv.org/abs/hep-th/9812238}{{\ttfamily arXiv:hep-th/9812238
  [hep-th]}}.

\bibitem{Rashkov:1999ji}
R.~C. Rashkov, ``{Note on the boundary terms in AdS / CFT correspondence for
  Rarita-Schwinger field},''
  \href{http://dx.doi.org/10.1142/S0217732399001887}{{\em Mod. Phys. Lett.}
  {\bfseries A14} (1999) 1783--1796},
\href{http://arxiv.org/abs/hep-th/9904098}{{\ttfamily arXiv:hep-th/9904098
  [hep-th]}}.

\bibitem{Matlock:1999fy}
P.~Matlock and K.~S. Viswanathan, ``{The AdS / CFT correspondence for the
  massive Rarita-Schwinger field},''
  \href{http://dx.doi.org/10.1103/PhysRevD.61.026002}{{\em Phys. Rev.}
  {\bfseries D61} (2000) 026002},
\href{http://arxiv.org/abs/hep-th/9906077}{{\ttfamily arXiv:hep-th/9906077
  [hep-th]}}.

\bibitem{Policastro:2008cx}
G.~Policastro, ``{Supersymmetric hydrodynamics from the AdS/CFT
  correspondence},''
  \href{http://dx.doi.org/10.1088/1126-6708/2009/02/034}{{\em JHEP} {\bfseries
  02} (2009) 034},
\href{http://arxiv.org/abs/0812.0992}{{\ttfamily arXiv:0812.0992 [hep-th]}}.

\bibitem{Gauntlett:2011mf}
J.~P. Gauntlett, J.~Sonner, and D.~Waldram, ``{Universal fermionic spectral
  functions from string theory},''
  \href{http://dx.doi.org/10.1103/PhysRevLett.107.241601}{{\em Phys. Rev.
  Lett.} {\bfseries 107} (2011) 241601},
\href{http://arxiv.org/abs/1106.4694}{{\ttfamily arXiv:1106.4694 [hep-th]}}.

\bibitem{Gauntlett:2011wm}
J.~P. Gauntlett, J.~Sonner, and D.~Waldram, ``{Spectral function of the
  supersymmetry current},''
  \href{http://dx.doi.org/10.1007/JHEP11(2011)153}{{\em JHEP} {\bfseries 11}
  (2011) 153},
\href{http://arxiv.org/abs/1108.1205}{{\ttfamily arXiv:1108.1205 [hep-th]}}.

\bibitem{Erdmenger:2013thg}
J.~Erdmenger and S.~Steinfurt, ``{A universal fermionic analogue of the shear
  viscosity},'' \href{http://dx.doi.org/10.1007/JHEP07(2013)018}{{\em JHEP}
  {\bfseries 07} (2013) 018},
\href{http://arxiv.org/abs/1302.1869}{{\ttfamily arXiv:1302.1869 [hep-th]}}.

\bibitem{Liu:2013fja}
J.~T. Liu, L.~A. Pando~Zayas, and Z.~Yang, ``{Small Treatise on Spin-3/2 Fields
  and their Dual Spectral Functions},''
  \href{http://dx.doi.org/10.1007/JHEP02(2014)095}{{\em JHEP} {\bfseries 02}
  (2014) 095},
\href{http://arxiv.org/abs/1401.0008}{{\ttfamily arXiv:1401.0008 [hep-th]}}.

\bibitem{Deser:1977uq}
S.~Deser and B.~Zumino, ``{Broken Supersymmetry and Supergravity},''
\href{http://dx.doi.org/10.1103/PhysRevLett.38.1433}{{\em Phys. Rev. Lett.}
  {\bfseries 38} (1977) 1433--1436}.

\bibitem{Denef:2009yy}
F.~Denef, S.~A. Hartnoll, and S.~Sachdev, ``{Quantum oscillations and black
  hole ringing},'' \href{http://dx.doi.org/10.1103/PhysRevD.80.126016}{{\em
  Phys. Rev.} {\bfseries D80} (2009) 126016},
\href{http://arxiv.org/abs/0908.1788}{{\ttfamily arXiv:0908.1788 [hep-th]}}.

\bibitem{Haehl:2018izb}
F.~M. Haehl and M.~Rozali, ``{Effective Field Theory for Chaotic CFTs},''
  \href{http://dx.doi.org/10.1007/JHEP10(2018)118}{{\em JHEP} {\bfseries 10}
  (2018) 118},
\href{http://arxiv.org/abs/1808.02898}{{\ttfamily arXiv:1808.02898 [hep-th]}}.

\bibitem{Haehl:2019eae}
F.~M. Haehl, W.~Reeves, and M.~Rozali, ``{Reparametrization modes, shadow
  operators, and quantum chaos in higher-dimensional CFTs},''
  \href{http://dx.doi.org/10.1007/JHEP11(2019)102}{{\em JHEP} {\bfseries 11}
  (2019) 102}, \href{http://arxiv.org/abs/1909.05847}{{\ttfamily
  arXiv:1909.05847 [hep-th]}}.

\end{thebibliography}\endgroup

\end{document}